\documentclass[
    aps,
    prl,
    reprint,
    superscriptaddress
]{revtex4-2}

\usepackage{graphicx}
\usepackage{amsmath}
\usepackage{amssymb}
\usepackage{bm}

\begin{document}
\title{Gas Beam Dump and Power Meter for High Energy Lasers}

\author{H. Rajesh}
\email{hrajesh@stanford.edu}
\affiliation{Department of Mechanical Engineering, Stanford University,
Stanford, California 94305, USA}

\author{S. Cao}
\affiliation{Department of Mechanical Engineering, Stanford University,
Stanford, California 94305, USA}

\author{K. Ou}
\affiliation{Department of Mechanical Engineering, Stanford University,
Stanford, California 94305, USA}

\author{D. Singh}
\affiliation{Department of Mechanical Engineering, Stanford University,
Stanford, California 94305, USA}

\author{P. Michel}
\affiliation{Lawrence Livermore National Laboratory,
Livermore, California 94550, USA}

\author{M. R. Edwards}
\email{mredwards@stanford.edu}
\affiliation{Department of Mechanical Engineering, Stanford University,
Stanford, California 94305, USA}

\begin{abstract}
Gaseous optics exhibit substantially higher laser-induced-damage thresholds than solid materials, enabling high-intensity operation without permanent degradation. In this Letter, we describe a gas-based beam dump and power meter that exploits the strong absorption of ozone in the Hartley band (200–320 nm). We develop and experimentally validate a model for ozone absorption that predicts the propagation length required for complete energy deposition and the resulting gas temperature rise. By calibrating the temperature rise against incident pulse energy, we demonstrate accurate millijoule-scale pulse energy measurements with an all-gas absorber. These results establish gas-phase absorbers as a scalable, debris-free platform for high-damage-threshold beam dumps and energy diagnostics for high-fluence lasers.
\end{abstract}

\maketitle

High-energy lasers delivering kilojoule- to megajoule-scale pulses have enabled major advances in inertial confinement fusion, high-energy-density physics, and laser-driven radiation sources \cite{NIF,AbuShawareb2022lawson,drake2006hedp,MichelLaser,Remington2006}. However, these facilities also generate substantial stray laser energy, including residual fundamental light after frequency conversion, back-reflections from optical components, and scattering from particles, debris, or structural elements within the beamline. This unwanted energy is typically dissipated in beam dumps based on absorbing solid surfaces. At high average powers, such dumps can be actively water cooled and used as calorimeters by calibrating the water temperature rise against the incident pulse energy~\cite{Williams2018Flowing}. However, solid beam dumps remain susceptible to damage from nanosecond pulses at fluences of tens of $\mathrm{J/cm^2}$ \cite{Zheng2022design}, where rapid heating, thermal stress, and dielectric breakdown become difficult to avoid. This constraint is increasingly restrictive for facility-scale high-energy lasers, often requiring substantial beam expansion, large standoff distances, or complex dump geometries to keep fluences below damage thresholds.

\begin{figure}
    \centering
    \includegraphics[width=\linewidth]{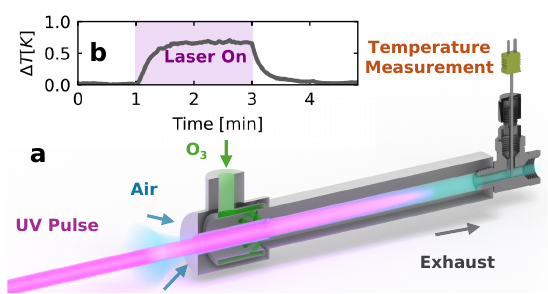}
    \caption{(a) Schematic showing an ozone power meter. A UV beam is absorbed in a flowing channel of ozone and air which is subsequently heated. The temperature rise of the gas can be calibrated to determine pulse energy. (b) Measured temperature rise for an 8 mJ 266 nm pulse at 10 Hz ($\dot Q_{\mathrm{total}} = 6~ \mathrm{LPM}, \mathrm{X_{O_3}} \approx 1\%$).}
    \label{fig:energymeter}
\end{figure}

In this work, we propose a gas-based beam dump and energy meter for nanosecond ultraviolet pulses, with an anticipated damage threshold more than two orders of magnitude higher than that of existing solid-state devices. The device, shown in Fig.~\ref{fig:energymeter}, uses the strong Hartley-band absorption of ozone, spanning approximately 200-320 nm, to dissipate the energy of ultraviolet (UV) laser pulses \cite{Wayne2000Chemistry}. The pulse energy is then calculated by calibrating the resulting temperature rise of the gas. The maximum tolerable fluence is expected to be set by dielectric breakdown of the gas and the onset of ionization-induced backscatter. Recent experiments with ozone--based diffraction gratings suggest that this ionization-based damage threshold exceeds $1~\mathrm{kJ/cm^2}$ for nanosecond pulses \cite{Michine2020Ultra}. Because the absorbing medium is continuously refreshed, the device is self-healing and reduces the need for replacement or repair. The use of a gaseous interaction medium further eliminates debris generation, a key limitation of solid-state devices.

Consider a laser pulse with fluence \(U\) propagating through a uniform absorbing gas with saturation fluence \(U_s=\hbar\omega/\sigma\), where \(\omega\) is the laser frequency, \(\sigma\) is the gas absorption cross section at \(\omega\), and \(\hbar\) is the reduced Planck constant. Generally, the evolution of a laser pulse in an absorber is described by the change in intensity with time and propagation distance \(z\). However, for ozone photodissociation by nanosecond pulses, absorption is governed by the time-integrated photon exposure rather than the instantaneous intensity, since each absorbed photon can dissociate an ozone molecule and remove it from the absorbing population during the pulse.

Therefore, the change of $u(z) = U(z)/U_s$ over a propagation distance $z$ where the absorbing molecule has density $n_a$ is \cite{Ou2024high}:
\begin{equation}
\frac{du}{dz} = \sigma n_a \left(e^{-u} - 1 \right).
\label{eq:absorption}
\end{equation}
For a laser pulse with intial fluence $u_0$, the solution is
\begin{equation}
u(z) = \ln \left[ 1+e^{-\sigma n_a z} \left(e^{u_0}-1 \right) \right].
\label{eq:absorption_full}
\end{equation}
In the highly saturated limit, where $u_0-\sigma n_a z \gg 1$ Eq.~\ref{eq:absorption_full} reduces to
\begin{equation}
    u(z) \simeq u_0 - \sigma n_a z.
\end{equation}
At the beam-dump entrance where $z=0$, this condition becomes $u_0 \gg 1$.

In this regime, the propagation length required to entirely absorb the laser pulse ($L$) is then
\begin{equation}
    L = \frac{u_0}{\sigma n_a} = \frac{U_0}{\hbar \omega n_a} = \frac{U_0}{\hbar \omega X_a n},
    \label{eq:length1}
\end{equation}
where $X_a$ is the mole fraction of the absorbing species and $n$ is the total gas number density. 

 The condition $u_0 \gg 1$ is generally satisfied for the strongly absorbing gases and high laser fluences for which a gas beam dump would be most suitable. The saturation fluence of ozone at $\lambda = 266$ nm ($U_s =~$ 77 mJ/cm$^2$) is much less than the $>1$ kJ/cm$^2$ fluence at which a gas device could operate. As the laser pulse propagates through the beam dump, the distance $z$ will increase until $\sigma n_a z \approx u_0$. Therefore, for $u$ to reach 0 (complete absorption), we  necessarily violate the condition $u_0 - \sigma n_a z \gg 1$ and thus invalidate Eq.~\ref{eq:length1}. However, for the high-fluence pulses considered here, most of the propagation through the gas beam dump occurs in this high-fluence regime. The residual low-fluence tail is absorbed over only a few linear absorption lengths, of order $\sim 1/(\sigma n_a)$, which is negligible compared with the total dump length L when the incident pulse fluence satisfies $u_0\gg 1$.

\begin{figure}
    \centering
    \includegraphics[width=\linewidth]{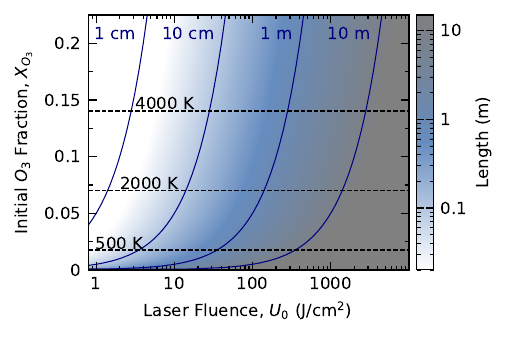}
    \caption{Required length for full absorption at varying imprint fluence and ozone concentration. Dashed lines show the expected temperature rise in the region of laser propagation.}
    \label{fig:lengthscaling}
\end{figure}

For ozone, the complete-absorption length can be written directly in terms of the total gas density $n$ and ozone mole fraction $X_{\mathrm{O}_3}$ as
\begin{equation}
    L^{[\mathrm{cm}]}
    \simeq
    1.34\times 10^{18}
    \frac{
    U_0^{[\mathrm{J/cm}^2]}
    }{
    X_{\mathrm{O}_3} n^{[\mathrm{cm}^{-3}]}
    },
    \label{eq:complete_absorption_length_numeric}
\end{equation}
for a $266~\mathrm{nm}$ laser pulse.
This length is plotted in Fig.~\ref{fig:lengthscaling} for an oxygen-ozone mixture (1 bar, 273K) as a function of the ozone fraction and initial beam fluence, showing, for example, that a 10\% mixture of ozone at standard temperature and pressure will completely absorb a 100 J/cm$^2$ laser pulse in 0.5 meters. The shortest beam dump for a fixed overall gas density requires the highest ozone concentration, but as we discuss below, high ozone concentrations may lead to unacceptably high gas temperatures when absorbing a high-energy beam.

\begin{figure}
    \centering
    \includegraphics[width=\linewidth]{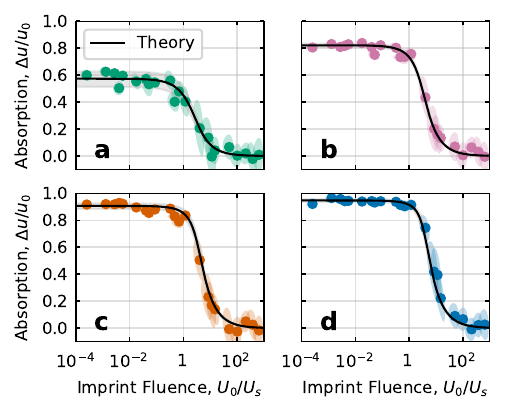}
    \caption{Experimental measurements of absorption at varying ozone concentrations ($X_{O_3}$) and ozone flow thickness ($l$). Points represent the average fluence and absorption across a selected portion of the beam while the ellipses show the spread. The black curves represent theory with the shaded gray bars representing uncertainty from concentration measurements.  (a) $X_{\mathrm{O_3}}$ = 1.40\%, $l$ = 2.5 mm (b) $X_{\mathrm{O_3}}$ = 4.0\%, $l$ = 2.5 mm (c) $X_{\mathrm{O_3}}$ = 5.0\%, $l$ = 2.5 mm (d) $X_{\mathrm{O_3}}$ = 2.5\%, $l$ = 5.0 mm }
    \label{fig:transmission_result}
\end{figure}

We validated the absorption model described in Eq.\ref{eq:absorption} experimentally using 266 nm, 5-10 ns pulses from a frequency quadrupled Nd:YAG laser focused through a flowing mixture of oxygen and ozone. Ozone was produced from oxygen at concentrations up to 5\% using a corona discharge generator. This flow had a height of 10 mm, a thickness varied between 2.5 and 5 mm, and a flow rate of 1 LPM. A co-flow of nitrogen surrounding the gas mixture was also introduced to limit shear-driven mixing with air.  A lens and CMOS camera were placed after the ozone flow to image the profile of the beam at the plane of the gas jet. Images were then collected with and without the ozone flow to measure absorption. 
Figure~\ref{fig:transmission_result} compares the measured absorption values to Eq. ~\ref{eq:absorption}. Each point represents the mean fluence and absorption and the shaded ellipses show the spread of both values over the region of interest. The measured values match Eq.\ref{eq:absorption} across a range of imprint fluences, ozone concentrations, and absorption lengths. At low imprint fluence $\left(U_0\ll U_s\right)$, the absorbed fraction remains constant, as expected in the linear regime where the number density of absorbers is large compared to the photon density. As the imprint fluence approaches and exceeds the saturation fluence $\left(U_0 \gg  U_s\right)$, the finite number of available absorbers begins to limit further absorption, reducing the absorbed fraction until it asymptotically approaches zero. We note that for absorption processes in which excited states of the absorbers relax back to the initial state on the timescale of the laser pulse, saturation is governed by the intensity, since absorbers may undergo multiple excitation–relaxation cycles during the pulse. In contrast, for photochemical dissociation of ozone in the Hartley band driven by a nanosecond laser pulse, the absorbing species is removed and does not reform within the pulse duration. Consequently, absorption is governed by the total number density of photons delivered.

Blocking a megajoule of laser energy with a gas requires the deposition of that energy in the gas, with the inevitable result that the gas will warm up. The limited heat capacity of gases means that the resulting heating can be substantial, and may set the practical limit on the minimum size of a gas-based beam dump. For $u\gg1$, we can assume that every ozone molecule absorbs a single photon and dissociates. Both the laser energy and the chemical energy released from the dissociation contribute to the heating of the gas. The final temperature of the gas in the region of laser propagation, once all modes have equilibrated, can be found from the equation:
\begin{equation}
\frac{U_0}{L}+U_{\mathrm{chem}} = \int_{T_0}^{T_f} c_v(T) n dT,
\label{eq:gasheating1}
\end{equation}
where $c_v$ is the average specific heat per molecule in the gas mixture and $U_{\mathrm{chem}}$ is the chemical energy released per unit volume. 

Numerical simulations of the reaction network in an ozone--oxygen mixture following UV photodissociation of ozone, using rate coefficients from Refs.~\cite{Burkholder2020, Green2000Kinetics, Michel2024photochemically}, show that when the initial ozone concentration exceeds 1\%, nearly all dissociated ozone is converted into ground-state oxygen within a few hundred microseconds. This result is consistent with experimental measurements of ozone concentration after UV energy deposition, probed using a low intensity UV beam. We therefore express the chemical energy released per dissociated ozone molecule as the enthalpy of formation of ozone, taking diatomic oxygen as the reference state:

\begin{equation}
U_{\mathrm{chem}} = n_{O_3} \Delta h_f^\circ \big( \mathrm{O_3}\big).
\label{eq:gasheating2}
\end{equation}

Here, $\Delta h_f^\circ \big( \mathrm{O_3}\big)$ is in units of energy per molecule. We note that for initial ozone concentrations below 1\%, a substantial fraction of the dissociation products recombine to reform ozone, driving the mixture back toward its initial ozone concentration rather than producing diatomic oxygen. However, such low concentrations are unsuitable for a beam dump due as the required length will be unreasonably long. As Eq.~\ref{eq:gasheating2} neglects ozone recombination, it will result in a slight overestimate in temperature. Assuming a constant specific heat, we can substitute Eqs.~\ref{eq:length1} and \ref{eq:gasheating2} into Eq.~\ref{eq:gasheating1} to determine the increase in temperature:

\begin{equation}
    \Delta T =
    \frac{
        \left[\hbar \omega + \Delta h_f^\circ\bigl(\mathrm{O}_3\bigr)\right]
        n_{\mathrm{O}_3}
    }{
        c_v n
    }
    =
    \frac{
        \hbar \omega + \Delta h_f^\circ\bigl(\mathrm{O}_3\bigr)
    }{
        c_v
    }
    X_{\mathrm{O}_3}.
\label{eq:gasheating3}
\end{equation}
Taking the heat capacity to be dominated by oxygen and including the chemical energy release, we can estimate the maximum equilibrium change in temperature due to a 266 nm pulse as:
\begin{equation}
\Delta T =  (28500 \textrm{ K}) \cdot X_{\textrm{O}_3}.
\label{eq:gasheating4}
\end{equation}
The availability of vibrational modes increases the heat capacity of oxygen at higher temperatures so we would expect the temperature rise to be somewhat less than this. This analysis also assumes the beam dump is just long enough to fully absorb the laser pulse. In practice, additional design margin increases the gas volume and further reduces the temperature rise. Figure \ref{fig:lengthscaling} shows the temperature rise for ozone--oxygen mixtures calculated using Eq.~\ref{eq:gasheating3}.

A beam dump used with repeating pulsed laser systems will require a constantly replenished volume of gas. For such a system to be viable, the gas must refresh such that each laser pulse interacts with a new volume. This requires imposing a condition on the volumetric flow rate of the input gas ($\dot Q_{\mathrm{gas}}$), which we can obtain by rewriting Eq.~\ref{eq:length1}:
\begin{equation}
\frac{\dot Q_{\mathrm{gas}}}{A} \geq \frac{U_0f_{\mathrm{rep}}}{\hbar \omega X_{\mathrm{O_3}}n},
\end{equation}
where $X_{\mathrm{O_3}}$ is the mole fraction, or concentration of ozone in the flow, $A$ is the area of the beam dump aperture, and $f_{\mathrm{rep}}$ is the repetition rate of the laser. At higher flow rates, only a fraction of the gas passes through the laser interaction region and absorbs energy. Assuming that this heated fraction mixes uniformly with the remainder of the flow downstream, the total energy deposited per laser pulse can be determined from the measured temperature increase of the output gas. As shown in Fig.~\ref{fig:energymeter}, the power meter uses a thermocouple to measure the gas temperature increase.

Assuming a constant $c_v$ and that all the energy goes into heating the gas, we can write the energy balance for this system:
\begin{equation}
(E_{\mathrm{laser}}+E_{\mathrm{chem}})f_{\mathrm{rep}} = \dot Q_{\mathrm{total}} c_v\Delta Tn, 
\end{equation}
where $E_{\mathrm{laser}}$ and $E_{\mathrm{chem}}$ are the energies deposited by a laser pulse and total chemical energy released per pulse respectively. $\dot Q_{\mathrm{total}} =\dot Q_{\mathrm{gas}}+\dot Q_{\mathrm{air}} $ is the total volumetric flow rate of gas including entrained air. Since the laser pulse energy determines the number of ozone molecules that are dissociated, we can write $E_{\mathrm{chem}}$ in terms of $E_{\mathrm{laser}}$.
\begin{equation}
E_{\mathrm{chem}} = \frac{\Delta h_f^\circ \big( \mathrm{O_3} \big) E_{\mathrm{laser}}}{\hbar \omega}.
\end{equation}
Now, we can relate the laser energy to the temperature rise as follows:

\begin{equation}
E_{\mathrm{laser}} =
\frac{
    \Delta T \, \dot{Q}_{\mathrm{total}} \, c_v \, n
}{
    f_{\mathrm{rep}}
    \left[
    1 + \frac{\Delta f_f^\circ(\mathrm{O}_3)}{\hbar \omega}
    \right]
}.
\label{eq:flowtemp1}
\end{equation}

Losses, including conductive and convective heat transfer, will depend on the geometry and material properties of the device and may be included in Eq.~\ref{eq:flowtemp1} as a coefficient. However, since these losses are approximately linear with respect to $\Delta T$, we can  account for them using a calibration factor $\eta$ and simplify Eq.~\ref{eq:flowtemp1} to
\begin{equation}
E_{\mathrm{laser}} = \frac{\eta \Delta T} {f_{\textrm{rep}}} .
\label{eq:simpleflowtemp1}
\end{equation}
As with solid-state energy meters, the factor $\eta $ is determined experimentally.

We built an ozone--based power meter like that shown in Fig.~\ref{fig:energymeter}. A 0.5 LPM mixture of oxygen with 2-4\% ozone flowed through a 15 cm-long additively manufactured resin tube and exhausted into a downstream vacuum pump. The pump was operated at 6 LPM, entraining ambient air through the meter’s 1 cm aperture. A thermocouple placed downstream measured the gas temperature.
\begin{figure}
    \centering
    \includegraphics[width=\linewidth]{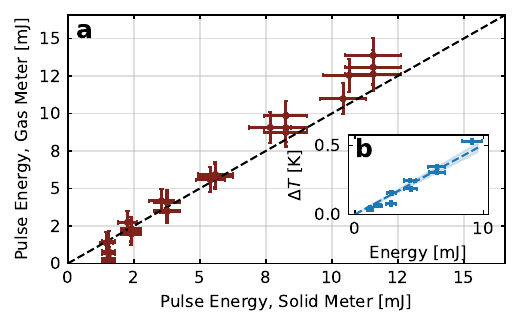}
    \caption{Experimental measurements of energy with the ozone energy meter. (a) Calibrated gas energy meter measurements compared to solid-state energy measurements. (b) Calibration of the gas energy meter. The measured temperature rise is compared to the pulse energy measured by a solid-state energy meter to obtain a calibration factor.}
    \label{fig:energyscaling}
\end{figure}
A 266 nm, 5-10 ns pulse was focused using a f = 500 mm lens. The pulse energy was adjusted over a range of 1 - 12 mJ, and measured using an Ophir UV energy meter. Trials were conducted at low fluence with a beam diameter of 7 mm ($U_{\mathrm{max}} \approx 30   ~\mathrm{mJ/cm^2}$) to determine the calibration coefficient. Following this, the beam diameter was reduced to 0.5 mm to increase the fluence ($U_{\mathrm{max}} \approx6~\mathrm{J/cm^2}$) for measurement trials. 

Figure~\ref{fig:energyscaling}a compares laser pulse energies measured with both the ozone and solid-state (UV) power meters, showing good agreement following calibration. The measured pulse energies are higher than what theory predicts. This is likely due to heat transfer losses, including convection and conduction to the channel walls. However, as shown, a linear calibration factor sufficiently accounts for these losses and results in calculated energies from the ozone energy meter that are generally within the error of the values measured with the solid-state energy meter. This calibration factor is specific to the operating conditions, environment, and energy meter geometry. If any of these parameters were to change drastically, recalibration would be required to ensure accurate observation. In the case of the results presented in Fig.~\ref{fig:energyscaling}, all measurements were conducted in a controlled laboratory environment with the same operating parameters. The measurements were conducted over the span of two days, indicating that the energy meter is robust against slight fluctuations in a standard laboratory environment. Although the range of energies tested was limited to the available laser energy, this demonstration serves as a proof of principle for a high-energy device.

The theory and experiments presented here provide the basis for designing gas beam dumps and power meters for high-energy UV lasers. Equation~\ref{eq:length1} prescribes required lengths based on achievable operating conditions and Eq.~\ref{eq:gasheating1} defines the maximum temperature in the device. Figure~\ref{fig:energymeter} shows a possible design for an energy meter. A high ozone concentration will reduce the required length but may increase the temperature beyond material limits. Increasing the aperture of the energy meter relative to the laser beam diameter may create a buffer of cooler gas surrounding the heated region, increasing the allowable temperature. Active cooling methods such as directed flows of air on the walls are simple methods to limit heating of the channel walls. 

While this work discusses the use of an ozone--based power meter for ultraviolet lasers, the concept presented here may be extended to other highly absorbing gases at different wavelengths. The general equations presented may be used to determine appropriate parameters. Per Eq.~\ref{eq:length1}, to achieve reasonable lengths (less than a few meters), absorption cross sections should be on the order of $10^{-18}$ $\mathrm{cm^2}$ or larger, although the achievable concentration also plays a role. For example, ozone produced from oxygen by corona discharge is typically limited to 20\% concentration. A non-exhaustive list of gas-wavelength combinations with appropriate cross sections may be found in Ref. \cite{Ou2024high}.

In conclusion, we present the theory required to design a gas beam dump and power meter and show experimental characterization of a prototype device. We experimentally validate a model describing absorption of ultraviolet light in ozone. Following calibration, experimental measurements from an ozone energy meter show agreement with a solid-state energy meter at millijoule energies, suggesting a new type of power meter that can be scaled to operate with high-energy laser systems.

\begin{acknowledgments}
This work was supported by the National Nuclear Security Administration
under Award Nos. DE-NA0004130 and DE-NA0004272; the U.S. National Science
Foundation under Award Nos. PHY-2308641 and PHY-2541940; and the Advanced
Research Projects Agency--Energy under Award No. DE-AR0002056.

This work was performed under the auspices of the U.S. Department of Energy
by Lawrence Livermore National Laboratory under Contract No.
DE-AC52-07NA27344 and was supported by the LLNL Laboratory Directed Research
and Development Program under Project No. 24-ERD-001.
\end{acknowledgments}

\section*{Disclosures}
Work at Stanford has been supported in part by Xcimer Energy LLC.

\section*{Data Availability}
Data underlying the results presented in this paper are not publicly
available at this time but may be obtained from the authors upon reasonable
request.

\bibliography{References}

@Article{Ou2024high,
  author = {K. Ou and V. M. Perez-Ramirez and S. Cao and C. Redshaw and J. Lee and M. M. Wang and J. M. Mikhailova and P. Michel and M. R. Edwards},
  title  = {High-energy transient gas pinholes via saturated absorption},
    journal   = {Optics Letters},
  year   = {2025},
  volume    = {50},
  doi       = {10.1364/OL.547141},
  issue     = 14,
  numpages  = {4},
  publisher = {Optica},
  url       = {https://opg.optica.org/ol/fulltext.cfm?uri=ol-50-4-1309},


}

@article{NIF,
  author       = {Spaeth, M. L. and Manes, K. R. and Kalantar, D. H. and Miller, P. E. and Heebner, J. E. and Bliss, E. S. and Spec, D. R. and Parham, T. G. and Whitman, P. K. and Wegner, P. J. and others},
  title        = {Description of the NIF Laser},
  doi          = {10.13182/FST15-144},
  url          = {https://www.osti.gov/biblio/1256427},
  journal      = {Fusion Science and Technology},
  issn         = {ISSN 1536-1055},
  number       = {1},
  volume       = {69},
  place        = {United States},
  publisher    = {American Nuclear Society},
  year         = {2017},
  month        = {03}}

@book{MichelLaser,
  author       = {Michel, P},
  title        = {Introduction to Laser-Plasma Interactions},
  doi          = {10.1007/978-3-031-23424-8},
  url          = {https://www.osti.gov/biblio/1970687},
  issn         = {ISSN 1868-4513},
  publisher = {Springer},
  place        = {United States},
  year         = {2023},
  month        = {01}}

@article{Remington2006,
  title = {Experimental astrophysics with high power lasers and $Z$ pinches},
  author = {Remington, Bruce A. and Drake, R. Paul and Ryutov, Dmitri D.},
  journal = {Rev. Mod. Phys.},
  volume = {78},
  issue = {3},
  pages = {755--807},
  numpages = {0},
  year = {2006},
  month = {Aug},
  publisher = {American Physical Society},
  doi = {10.1103/RevModPhys.78.755},
  url = {https://link.aps.org/doi/10.1103/RevModPhys.78.755}
}

@book{drake2006hedp,
  author    = {R. Paul Drake},
  title     = {High-Energy-Density Physics: Fundamentals, Inertial Fusion, and Experimental Astrophysics},
  publisher = {Springer},
  year      = {2006},

}

@Article{Michel2024photochemically,
  author    = {Michel, P. and Lancia, L. and Oudin, A. and Kur, E. and Riconda, C. and Ou, K. and Perez-Ramirez, V.M. and Lee, J. and Edwards, M. R.},
  journal   = {Physical Review Applied},
  title     = {Photochemically induced acousto-optics in gases},
  year      = {2024},
  month     = {Aug},
  pages     = {024014},
  volume    = {22},
  doi       = {10.1103/PhysRevApplied.22.024014},
  issue     = {2},
  numpages  = {14},
  publisher = {American Physical Society},
  url       = {https://link.aps.org/doi/10.1103/PhysRevApplied.22.024014},
}

@techreport{Burkholder2020,
  author = {Burkholder, J. and Sander, S. and Abbatt, J. and Barker, J. and Huie, R. and Kolb, C. and Kurylo, M. and Orkin, V. and Wilmouth, D. and Wine, P.},
  title = {Chemical Kinetics and Photochemical Data for Use in Atmospheric Studies: Evaluation Number 19},
  institution = {Jet Propulsion Laboratory, California Institute of Technology},
  year = {2020},
  number = {JPL Publication 19-5}
}

@Article{AbuShawareb2022lawson,
  author    = {Abu-Shawareb, H. and Acree, R. and Adams, P. and Adams, J. and Addis, B. and Aden, R. and Adrian, P. and Afeyan, B. B. and Aggleton, M. and Aghaian, L. and et al.},
  title     = {Lawson Criterion for Ignition Exceeded in an Inertial Fusion Experiment},
  journal   = {Physical Review Letters},
  year      = {2022},
  month     = {Aug},
  volume    = {129},
  pages     = {075001},
  doi       = {10.1103/PhysRevLett.129.075001},
  numpages  = {6},
  publisher = {American Physical Society},
  url       = {https://link.aps.org/doi/10.1103/PhysRevLett.129.075001},
}

@article{Zheng2022design,
title = {Design of laser beam dump with high laser-induced-damage threshold},
journal = {Optics and Laser Technology},
volume = {146},
pages = {107561},
year = {2022},
issn = {0030-3992},
doi = {https://doi.org/10.1016/j.optlastec.2021.107561},
url = {https://www.sciencedirect.com/science/article/pii/S0030399221006496},
author = {Tianran Zheng and Hongjie Liu and Fang Wang and Yong Xiang and Ye Tian and Zhufeng Shao and Yuan Chen and Dongxia Hu and Xiaodong Yuan},


}

@Book{Wayne2000Chemistry,
  author    = {Richard P. Wayne},
  title     = {Chemistry of Atmospheres: An Introduction to the Chemistry of the Atmospheres of Earth, the Planets, and Their Satellites},
  edition   = {3rd},
  publisher = {Oxford University Press},
  year      = {2000},
  isbn      = {9780198503750},
  url       = {https://global.oup.com/academic/product/chemistry-of-atmospheres-9780198503750}
}

@Article{Michine2020Ultra,
  author    = {Yurina Michine and Hitoki Yoneda},
  title     = {Ultra high damage threshold optics for high power lasers},
  journal   = {Communications Physics},
  volume    = {3},
  number    = {1},
  pages     = {24},
  year      = {2020},
  doi       = {10.1038/s42005-020-0286-6},
  url       = {https://www.nature.com/articles/s42005-020-0286-6}
}

@Article{Williams2018Flowing,
  author    = {Paul A. Williams and Joshua A. Hadler and Christopher L. Cromer and Xiaoyu X. Li and John H. Lehman},
  title     = {Flowing water optical power meter design, operation, and uncertainty for multi-kilowatt laser power measurements},
  journal   = {Metrologia},
  volume    = {55},
  number    = {3},
  pages     = {427–436},
  year      = {2018},
  doi       = {10.1088/1681-7575/aaae78},
  url       = {https://www.nist.gov/publications/flowing-water-optical-power-meter-design-operation-and-uncertainty-multi-kilowatt-laser}
}

@Article{Green2000Kinetics,
  author    = {Jack G. Green and Jichun Shi and John R. Barker},
  title     = {Photochemical kinetics of vibrationally excited ozone produced
in the 248 nm photolysis of o2/o3 mixtures},
  journal   = {The Journal
of Physical Chemistry A},
  volume    = {104},
  number = {26},
  pages     = {6218–6226},
  year      = {2000},
  doi       = {10.1021/jp000635k}
}

\end{document}